\documentclass[]{spie}  

\usepackage{amsmath,amsfonts,amssymb}
\usepackage{graphicx}
\usepackage[colorlinks=true, allcolors=blue]{hyperref}

\title{The In-Flight Noise Performance of the JWST/NIRSpec Detector System}

\author[a]{Stephan M. Birkmann}
\author[b]{Giovanna Giardino}
\author[a]{Marco Sirianni}
\author[c]{Pierre Ferruit}
\author[d]{Bernhard Rauscher}
\author[c]{Catarina Alves de Oliveira}
\author[a]{Torsten Böker}
\author[e]{Nimisha Kumari}
\author[a]{Nora Lützgendorf}
\author[e]{Elena Manjavacas}
\author[f]{Charles Proffitt}
\author[a]{Timothy D. Rawle}
\author[a]{Maurice te Plate}
\author[e]{Peter Zeidler}
\affil[a]{European Space Agency, STScI, Baltimore, USA}
\affil[b]{ATG Europe for ESA, ESTEC, Noordwijk, The Netherlands}
\affil[c]{European Space Agency, ESAC, Madrid, Spain}
\affil[d]{NASA Goddard Space Flight Center, Greenbelt, USA}
\affil[e]{AURA for ESA, STScI, Baltimore, USA}
\affil[f]{AURA, STScI, Baltimore, USA}

\authorinfo{Further author information: (Send correspondence to S.M.B)\\S.M.B.: E-mail: stephan.birkmann@esa.int}

\begin{document} 
\maketitle

\begin{abstract}
The Near-Infrared Spectrograph (NIRSpec) is one the four focal plane instruments on the James Webb Space Telescope (JWST) which was launched on December 25, 2021. We present the in-flight status and performance of NIRSpec’s detector system as derived from the instrument commissioning data as available at the time of the conference. The instrument features two 2048 × 2048 HAWAII-2RG sensor chip assemblies (SCAs) that are operated at a temperature of about 42.8 K and are read out via a pair of SIDECAR ASICs. NIRSpec supports “Improved Reference Sampling and Subtraction” (IRS2) readout mode that was designed to meet NIRSpec’s stringent noise requirements and to reduce 1/f and correlated noise. In addition, NIRSpec features subarrays optimized for bright object time series observations, e.g.\ for the observation of exoplanet transit around bright host stars. We focus on the dark signal as well as the read and total noise performance of the detectors. 
\end{abstract}

\keywords{JWST, NIRSpec, infrared detectors, noise}

\section{INTRODUCTION}
\label{sec:intro}  

\begin{figure}
    \centering
    \includegraphics{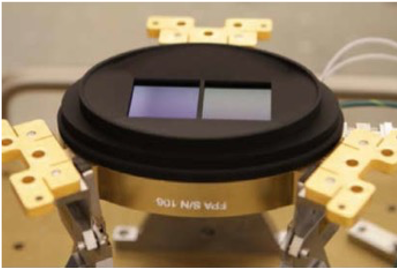}
    \caption{The NIRSpec Focal Plane Assembly (FPA), flight model \#106.}
    \label{fig:FPA}
\end{figure}

The Near-Infrared-Spectrograph (NIRSpec)\cite{Jakobsen2022} is one of the science instruments aboard the James Webb Space Telescope (JWST)\cite{gardner2006} launched on 25 December 2021. NIRSpec features an almost all reflective design with an optical bench manufactured out of Silicon-Carbide\cite{Salvignol2008}. It offers four different observing modes: 1) integral field spectroscopy (IFS)\cite{Boeker2022} via an image slicer\cite{Lobb2008}, 2) multi-object spectroscopy (MOS)\cite{Ferruit2022} via a micro-shutter array that employs about 250,000 individually addressable shutters\cite{Kutyrev2004}, 3) fixed slit (FS) spectroscopy via several long slits, and 4) bright object times series (BOTS)\cite{Birkmann2022} observations via a dedicated square aperture. In all observing modes the same selection of seven dispersers is available: one prism and six gratings selected via a grating wheel assembly\cite{Weidlich2006}, offering a wavelength coverage from $\sim$0.6 to 5.3$\;\mu m$ with varying spectral resolution. Light is detected by a pair of closely spaced HgCdTe HAWAII-2RG sensor chip assemblies (SCAs, see Fig.~\ref{fig:FPA})\cite{Rauscher_2014} with a nominal cut-off wavelength of $\sim$5.3$\;\mu m$. The detectors are operated at $T=42.8\;K$ and are read out by a pair of SIDECAR ASICs\cite{Loose2005}. Temperature drifts of the focal plane assembly (FPA) are kept to a minimum by means of active temperature control.

\subsection{NIRSpec Readout Modes} \label{sec:readout}

The NIRSpec detectors are read non-destructively ``up-the-ramp'' and offer two fundamentally different readout modes: the so-called traditional readout mode (TRAD) and the improved reference sampling and subtraction (IRS$^2$) readout mode. The latter is only supported for full-frame, and offers superior noise performance by means of sampling reference pixels regularly and interleaved with the science pixels, allowing for a better reference subtraction and thus lower correlated / 1/f noise\cite{Rauscher_2017}. Subarrays of different sizes are supported in traditional readout mode, using a slightly higher conversion gain compared to full frame mode. The higher conversion gain (e-/DN) is used in order to make the full physical well depth of the pixels accessible and support brighter targets before saturation, in particular important for BOTS observations.

For full-frame readout modes all pixels in the detector are read with a cadence or frame time $t_f$ of approximately 10.7 seconds (TRAD) or 14.8 seconds (IRS$^2$). The frame time for subarrays depend on the subarray size and for NIRSpec range from approximately 5.5 seconds for the largest subarray (ALLSLITS, covering all NIRSpec fixed slit apertures) down to $\sim$15\,ms for the smallest subarray (SUB32, used for target acquisition only). Readout modes that use all frames in an integration are called RAPID modes in the Astronomers Proposal Tool (APT), i.e.\ NRSRAPID for traditional readout mode and NRSIRS2RAPID for IRS$^2$ readout mode. In order to reduce the amount of memory needed in the on-board solid state recorder (SSR) for detector data, the NIRSpec readout modes are also offered in a frame averaged version, where multiple frames are averaged into groups on-board. For the traditional readout mode four frames are averaged into one group (NRS) and for IRS$^2$ five frames can be averaged into one group (NRSIRS2).

\section{NIRSpec Commissioning and Acquired Data}

The NIRSpec commissioning campaign started shortly after the successful launch of JWST on an Ariane 5 rocket with the power on and initialization of the Micro-shutter control electronics. After functional checkouts of the various NIRSpec sub-systems and the detector operating temperature had been reached, data acquisition including dark exposures started. A much more detailed description of the NIRSpec commissioning campaign and results is given at this conference\cite{BoekerSPIE, LuetzgendorfSPIE, GiardinoSPIE, AlvesSPIE, RawleSPIE}.

The analysis presented in this paper is based on the following set of dark exposures that were obtained between 04 March 2022 and 05 April 2022, after the NIRSpec FPA had reached its operating temperature of $T \sim 42.82\,$K and active temperature control had been enabled:
\begin{itemize}
    \item 54 traditional full frame dark exposures with 160 frames/groups each obtained in NRSRAPID mode
    \item 80 IRS$^2$ full frame dark exposures with 245 frames/groups each obtained in NRSIRS2RAPID mode
    \item 48 dark integrations with the ALLSLITS subarray with 265 frames/groups each in NRSRAPID mode
\end{itemize}
With the frame times listed in Sec.~\ref{sec:readout} it follows that the integration time for each of the traditional full frame darks was approximately 30 minutes, almost 60 minutes for the IRS$^2$ darks, and approximately 25 minutes for the ALLSLITS subarray darks.

\section{Data Reduction}
The ``ramps-to-slopes'' data processing was performed using the pre-processing pipeline developed for ground test and commissioning purposes by the ESA NIRSpec Science Operations Team. This pipeline is also used to implement and test algorithms that are then used to inform the development of the official JWST data processing pipeline run by STScI. The following processing steps were performed:
\begin{itemize}
    \item saturation detection and flagging
    \item mast bias subtraction
    \item reference pixel subtraction / correction, in the case of IRS$^2$ data including the improved data processing using the additionally available reference pixels and reference output
    \item non-linearity correction
    \item slope estimation using optimum weights\cite{Fixsen2000}, including jump detection (cosmic ray mitigation) using 2-point differences\cite{Anderson2011}, including multiplication with the conversion gain in order to go from DN/s to e$^-$/s
\end{itemize}

The above yielded a 2D electron rate map for each integration / exposure. The pixel-to-pixel median of these maps was computed to derive the dark signal maps for the different readout modes (see Sec.~\ref{sec:dark}).

We also computed the correlated double sample (CDS) for all frame pairs in all integrations / exposures. The CDS maps were then averaged using sigma clipping (3 iterations with a clipping threshold of 3 sigma) and the resulting standard deviation (for each pixel) is the CDS noise, which is $\sqrt{2}$ times the single frame read noise. CDS noise results are presented in Sec.~\ref{sec:cds}.

Finally, we repeated the slope estimation for all input data, limiting the used number of groups $n_g$ in the ramp to 5, 10, 15, and so on, up to the maximum available number of groups for that readout mode. For each of these sets of electron rate maps we computed the mean and standard deviation $\sigma(i, j, n_g)$ for each pixel ($i, j$) as a function of used groups $n_g$ using sigma clipping (3 iterations with a clipping threshold of 3 sigma). For the RAPID readout modes ($n_f = 1$ and thus $t_g = t_f$) the total noise $\sigma_{total}(i, j, n_g)$ is then:
\begin{equation}
    \sigma_{total}(i, j, n_g) = \sigma(i, j, n_g) \times t_{eff}(n_g) = \sigma(i, j, n_g) \times (n_g - 1) \times t_f,
\end{equation}
where $t_{eff}(n_g)$ is the effective integration time and $t_f$ is the frame time. The total noise as a function of effective integration time / number of groups is reported in Sec.~\ref{sec:noise}.

\section{Results} \label{sec:results}

The commissioning results for dark signal, CDS and total noise are summarized in the following sections.

\subsection{Dark Signal} \label{sec:dark}

\begin{figure}
    \centering
    \includegraphics[height=0.48\textheight]{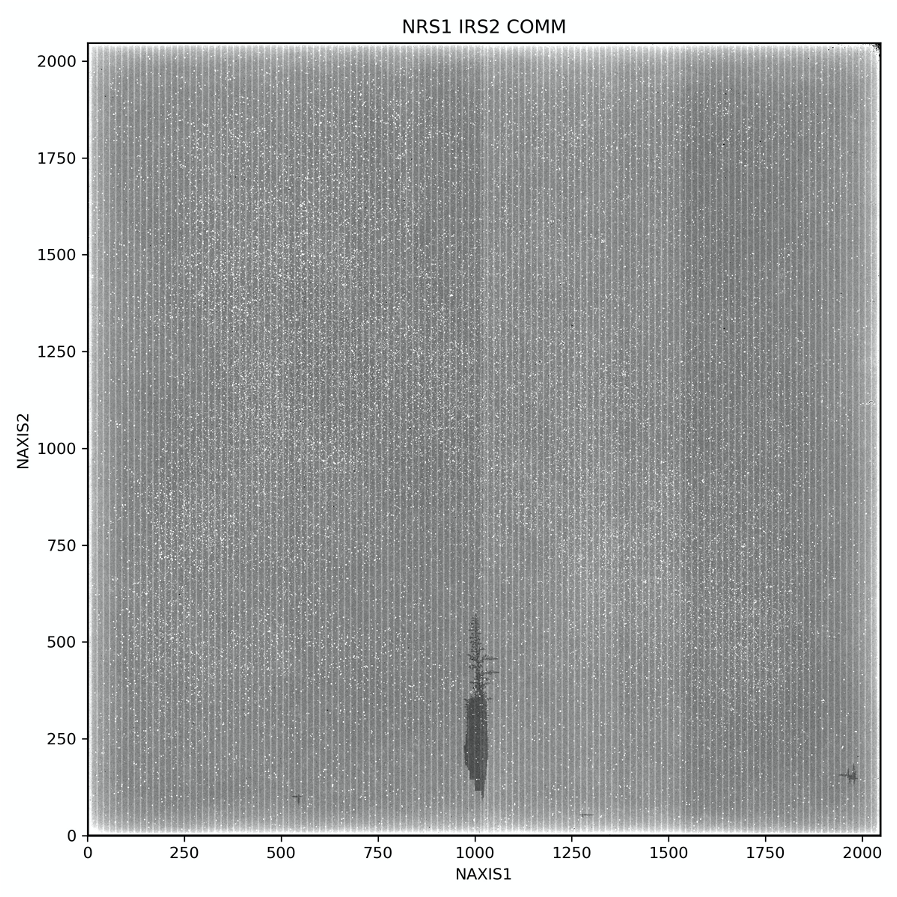}
    \includegraphics[height=0.48\textheight]{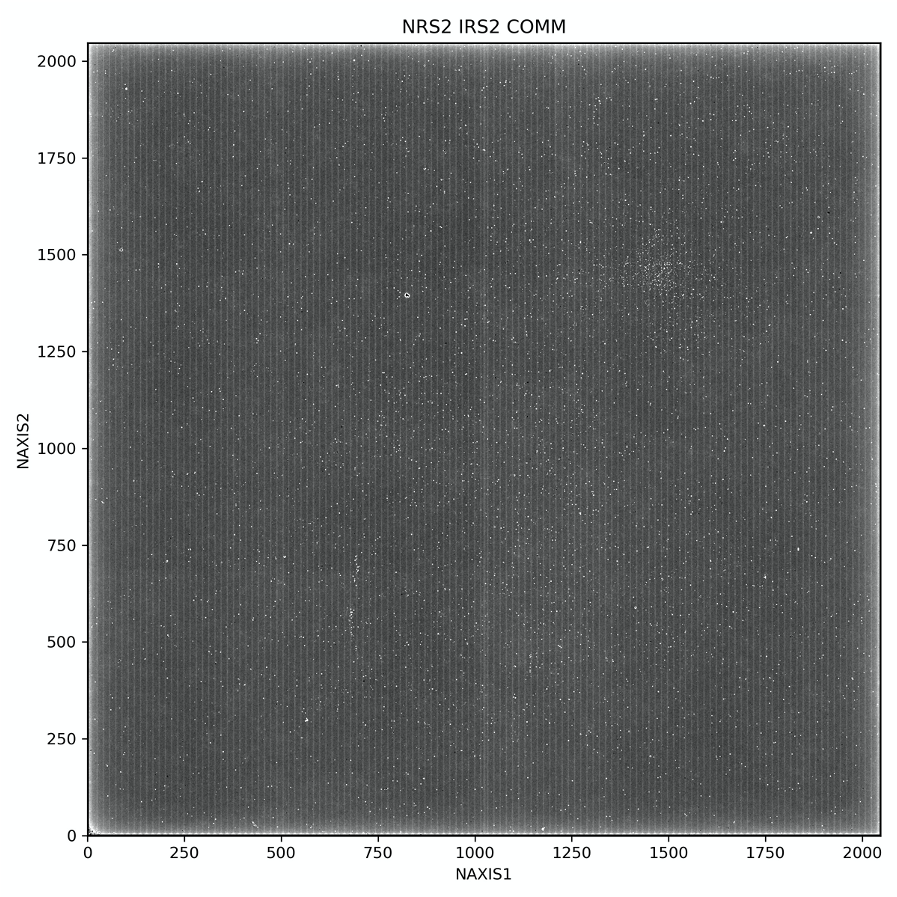}
    \caption{Dark signal maps for NRS1 (top) and NRS2 (bottom) for IRS$^2$ readout mode as obtained during JWST/NIRSpec commissioning. Linear grey scale stretch from 0 to 0.015 e$^-$/s.}
    \label{fig:dark}
\end{figure}

The median dark signal of the two NIRSpec SCAs for the different readout modes is presented in Table~\ref{tab:dark}. Figure~\ref{fig:dark} shows the dark signal maps for IRS$^2$ readout mode for both detectors.

\begin{table}[h]
    \centering
    \begin{tabular}{c|c|c|c}
         &  \multicolumn{3}{c}{Readout mode}\\
         Detector & TRAD & IRS2 & ALLSLITS\\\hline
         NSR1 & 0.0090 (0.0077) & 0.0082 (0.0071) & 0.0223 (0.0136)\\
         NRS2 & 0.0069 (0.0051) & 0.0048 (0.0040) & 0.0153 (0.0137)\\
    \end{tabular}
    \caption{Median dark signal in e$^-$/s for the NIRSpec detectors for traditional full frame (TRAD), IRS$^2$ (IRS2), and ALLSLITS subarray (ALLSLITS) readout modes as measured during commissioning. Comparison numbers from the last ground test are in brackets.}
    \label{tab:dark}
\end{table}

As known from ground testing, the dark signal for NRS1 is higher than for NRS2. The region of lower dark signal in the bottom center of NRS1
is due to a void in the epoxy back-filling between the detector and its attached readout integrated circuit (ROIC). The dark signal is higher for readout modes with shorter frame times, as most of the dark signal is not due to dark current, but rather multiplexer glow that occurs during readout\cite{Regan2000}. Other differences are due to the different detector tuning (supply voltages and currents) that are unique to each detector and readout mode.

However, the dark signal is higher than measured during the last cryogenic ground test in 2017\cite{Kimble2018} for both detectors, with a pronounced increase towards the edges of the arrays. This increase is likely related to the cosmic ray environment at L2, see discussion in Sec.~\ref{sec:cr}.

Even with the small increase compared to ground, the dark signal is still very low for most pixels and not a driver for the total noise.

\subsection{CDS Noise} \label{sec:cds}

The correlated double sample noise measured during commissioning is presented in Table~\ref{tab:cds}. It is in line with the numbers derived during the last cryogenic ground test, also indicating that the detector tuning is unchanged and as expected.

\begin{table}[h]
    \centering
    \begin{tabular}{c|c|c|c}
        &  \multicolumn{3}{c}{Readout mode}\\
         Detector & TRAD & IRS2 & ALLSLITS\\\hline
         NRS1 & 13.0 (12.9) & 9.59 (9.61) & 11.2 (11.4) \\
         NRS2 & 13.1 (13.1) & 11.6 (11.6) & 10.7 (11.0) \\
    \end{tabular}
    \caption{Median CDS noise in data numbers (DN) for the NIRSpec detectors for traditional full frame (TRAD), IRS$^2$ (IRS2), and ALLSLITS subarray (ALLSLITS) readout modes as measured during commissioning. Comparison numbers from the last ground test are in brackets.}
    \label{tab:cds}
\end{table}

As the reported CDS noise numbers are in DN they appear lower for the ALLSLITS subarray than the traditional full frame, because of the difference in conversion gain between the two readout modes (conversion gain in e$^-$/DN is about a factor 1.4 higher for subarrays than that used for full frame).

\begin{figure}
    \centering
    \includegraphics[height=0.48\textheight]{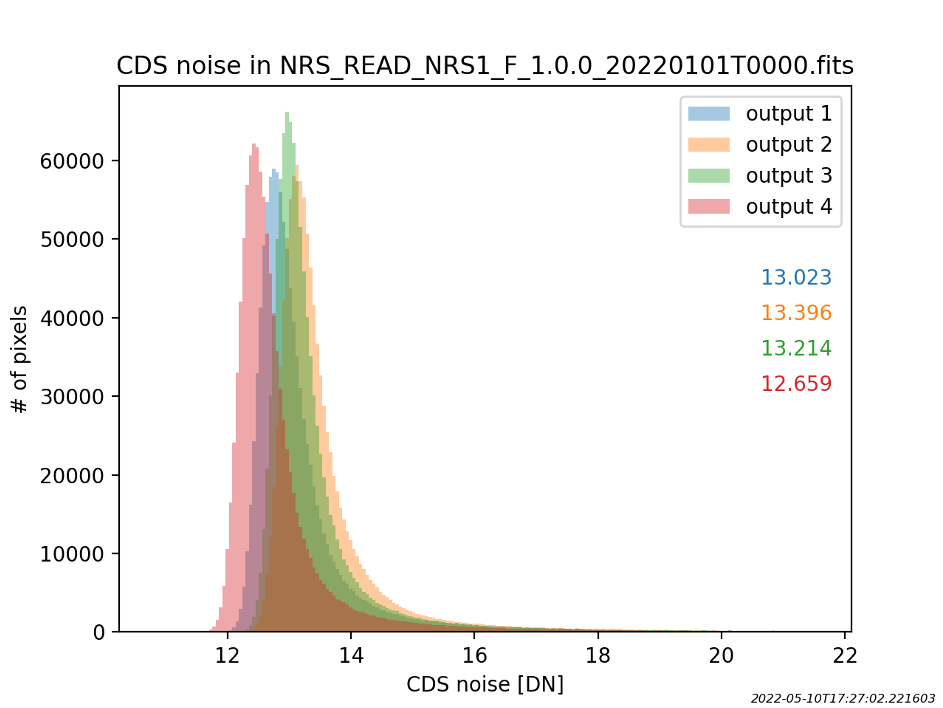}
    \includegraphics[height=0.48\textheight]{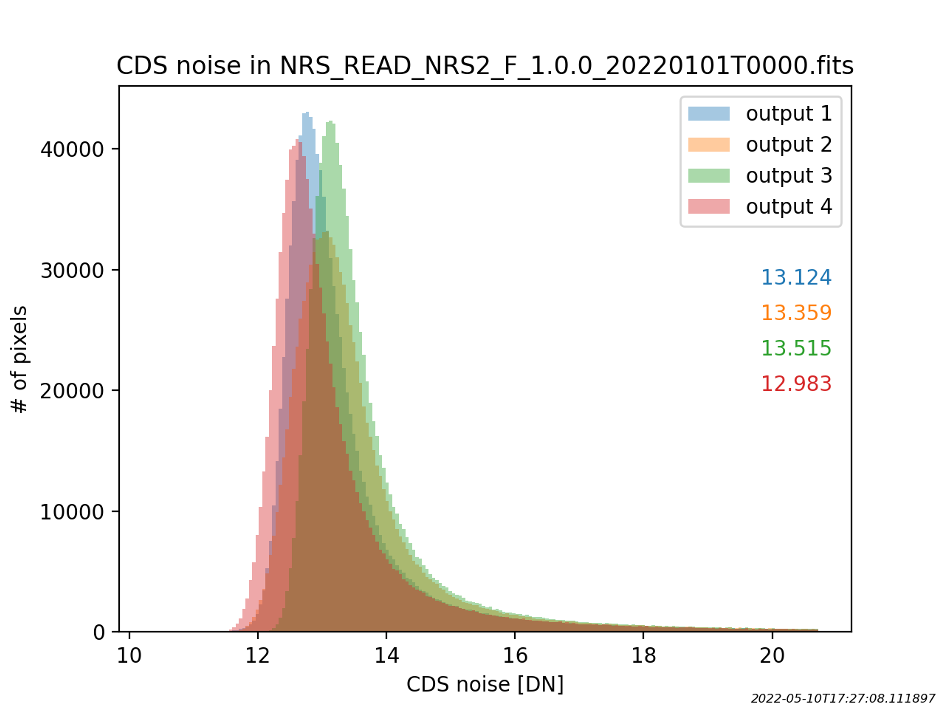}
    \caption{CDS noise histograms for the traditional full frame readout mode for NRS1 (top) and NRS2 (bottom) as measured during commissioning.}
    \label{fig:cds}
\end{figure}

The numbers reported in Table~\ref{tab:cds} are the median for the full detector. There are small differences between the outputs of the detectors, as is illustrated by the histograms in Fig.~\ref{fig:cds}.

\subsection{Total Noise} \label{sec:noise}

\begin{figure}
    \centering
    \includegraphics[width=0.86\textwidth]{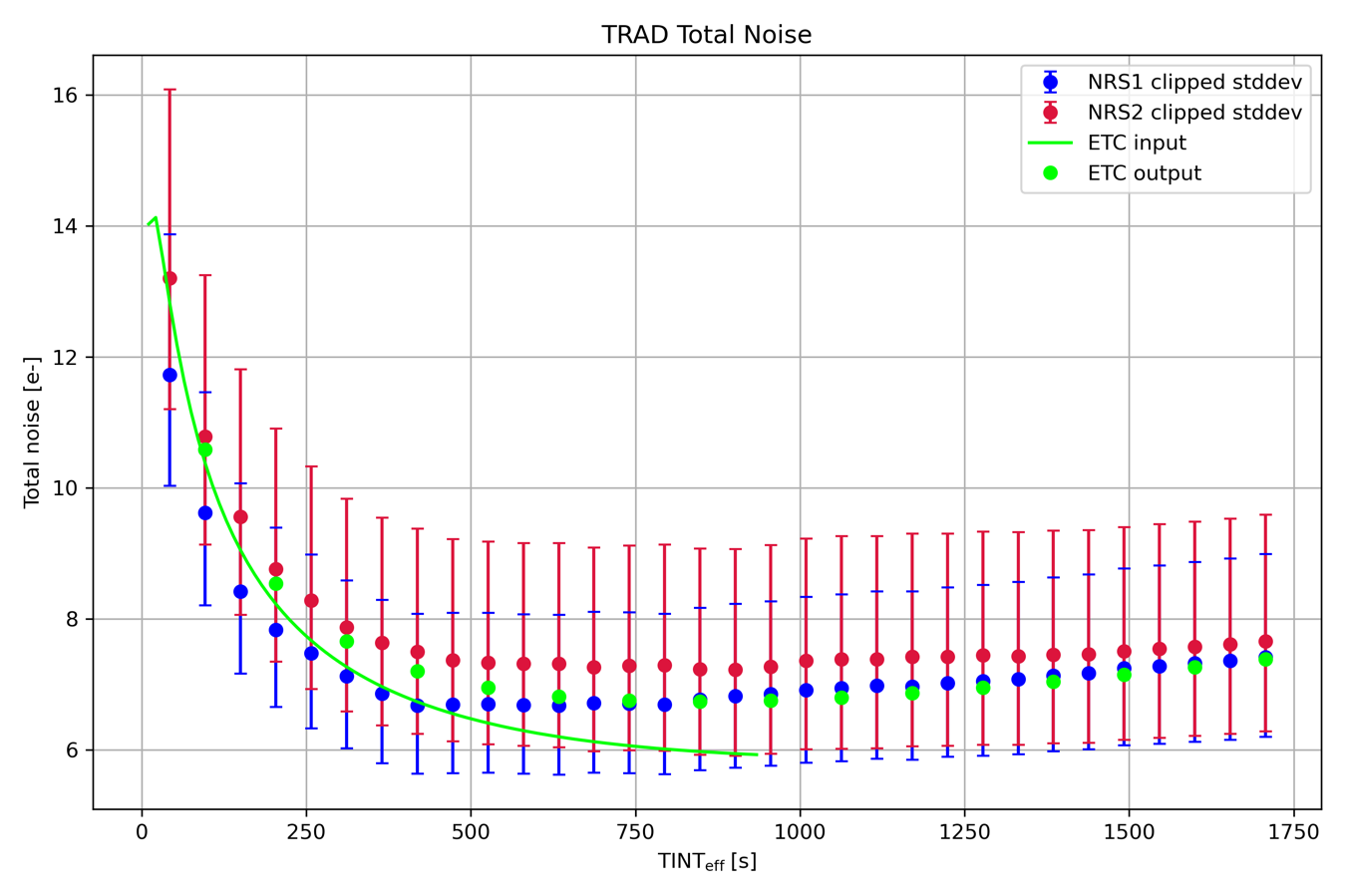}
    \caption{Total noise in electrons as a function of effective integration time for traditional full frame readout mode. Blue and red symbols are for median of the two detectors NRS1 and NRS2, respectively, the denoted error bars show the 10 and 90 percentiles. Green symbols are for the total noise as reported by the ETC. The green line is the medium total noise averaged for both detectors as measured on ground.}
    \label{fig:noise:trad}
\end{figure}

\begin{figure}
    \centering
    \includegraphics[width=0.86\textwidth]{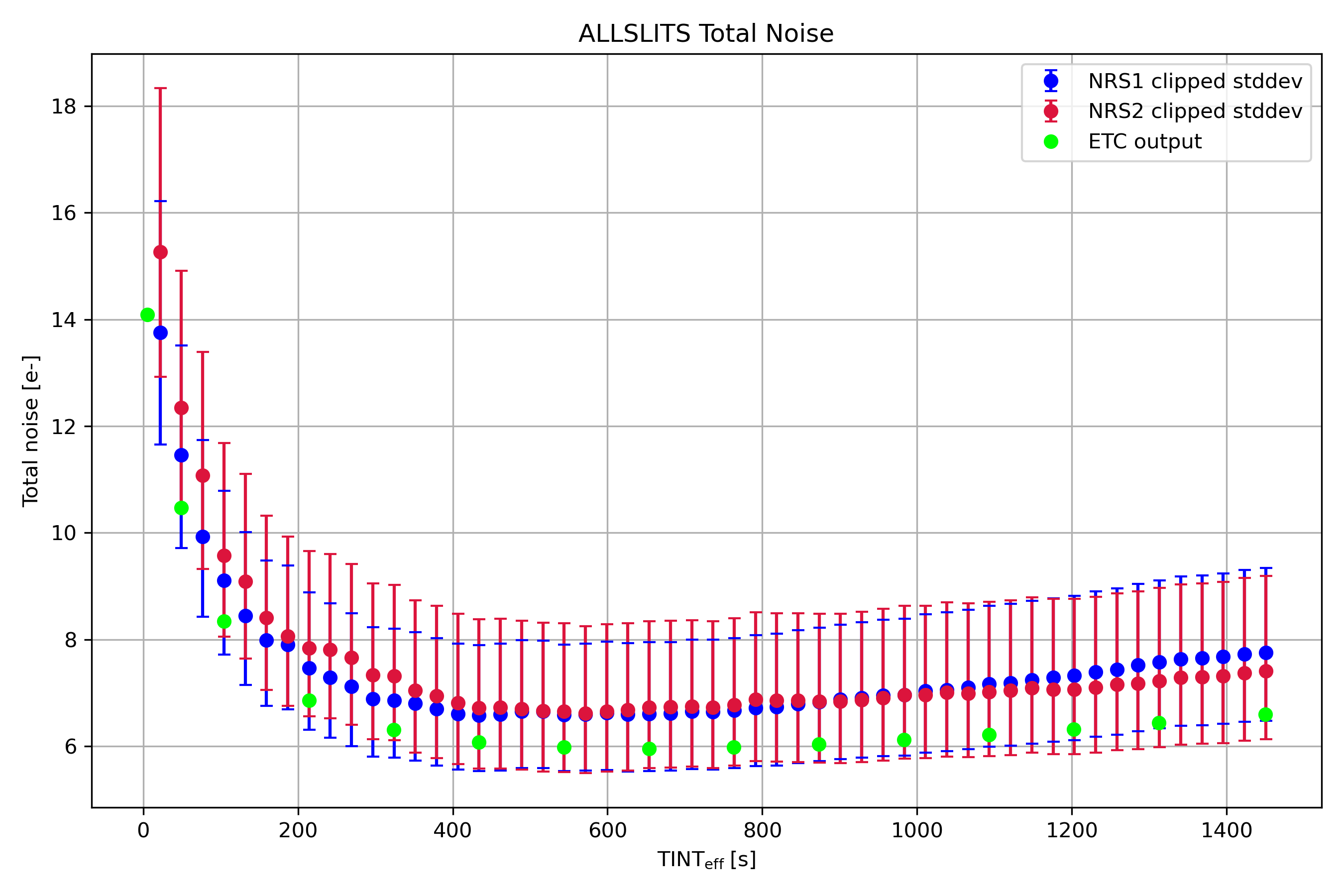}
    \caption{Total noise in electrons as a function of effective integration time for IRS$^2$ readout mode. Otherwise like Fig.~\ref{fig:noise:trad}.}
    \label{fig:noise:irs2}
\end{figure}

\begin{figure}
    \centering
    \includegraphics[width=0.86\textwidth]{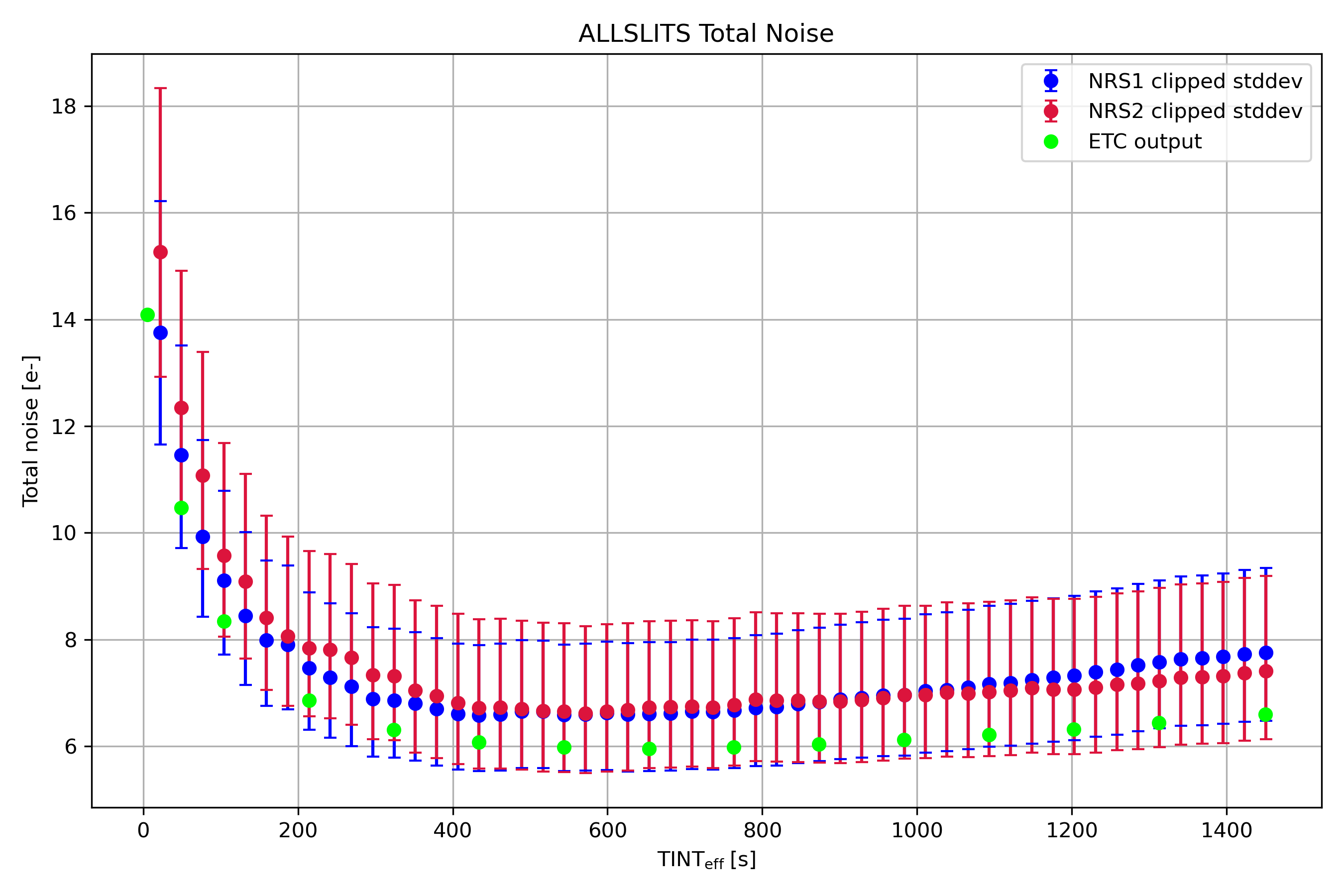}
    \caption{Total noise in electrons as a function of effective integration time for the ALLSLITS subarray. Otherwise like Fig.~\ref{fig:noise:trad}.}
    \label{fig:noise:sub}
\end{figure}

The total noise as a function if effective integration time / ramp length is shown in Figures~\ref{fig:noise:trad} through \ref{fig:noise:sub} for the traditional full frame, IRS$^2$, and ALLSLITS subarray readout modes, respectively, and is summarized in Table~\ref{tab:noise} below.

\begin{table}[h]
    \centering
    \begin{tabular}{c|c|c|c}
        & \multicolumn{3}{c}{Total noise [e$^-$]} \\
         Readout / Detector & T$_{eff}\sim$950\,s & $\sim$1700\,s & $\sim$3560 \\\hline\hline
         TRAD / NRS1 & 6.9 & 7.4 & N/A \\
         TRAD / NRS2 & 7.3 & 7.7 & N/A \\\hline
         IRS2 / NRS1 & 5.9 & 6.6 & 8.5 \\
         IRS2 / NRS2 & 7.2 & 7.6 & 9.2 \\\hline
         SUB / NRS1 & 7.0 & 7.8 & N/A \\
         SUB / NRS2 & 7.0 & 7.5 & N/A \\
    \end{tabular}
    \caption{Total noise for the two NIRSpec detectors for different readout modes and effective integration times as measured during commissioning.}
    \label{tab:noise}
\end{table}

The total noise is higher than measured during ground testing\cite{Birkmann2018}. This is expected and can be attributed to the cosmic ray environment at L2. Cosmic rays have the following effects on data that results in an increase of total noise:
\begin{itemize}
    \item Detected cosmic ray hits are flagged and the ramp is broken into multiple segments, resulting in a higher noise of this integration compared an undisturbed one.
    \item Some cosmic ray hits are strong enough to saturate one or more pixels, resulting in the loss of data after the hit occurred, shortening the effective integration time and increasing the noise.
    \item Residuals of undetected (jump below detection threshold) cosmic rays.
    \item Other secondary effects, like the charge trapping (``inverse persistance'') that will occur after a significant jump in a detector pixel.
\end{itemize}

As is evident from the plots, the average total noise is in line with the predictions / noise model used by the JWST exposure time calculator (ETC)\cite{Pickering2016} for the two full frame readout modes, traditional and IRS$^2$. The ETC underestimates the total noise for the ALLSITS subarray, which is likely due to limitations in the internal noise model, where the same readout noise is assumed for traditional and subarray readout mode.

For all readout modes, total noise decreases with increasing integration times to approximately 500 seconds where it levels out and then increases with longer integration times. For detector noise limited observations of faint sources it is still beneficial to use long integration times for optimum signal-to-noise ratios. However, due to potential early saturation after a strong cosmic ray hit, it is advisable to have multiple integrations per observation, ideally in the form of dithered exposures.

\subsection{Cosmic rays} \label{sec:cr}

As discussed above, the slight increase in measured total noise can be attributed to the impacts of cosmic rays. The cosmic ray rate is in line with pre-flight predictions\cite{Giardino2019}, with approximately 60\% of all pixels being affected by one or more cosmic ray events in a one hour exposure. In our one hour long darks, only 2 to 3\% of pixels saturated prematurely due to strong cosmic ray events. Due to inter-pixel capacitive (IPC) coupling\cite{Moore2004}, typical cosmic rays affect at least 5 pixels, i.e. one cosmic ray leads to a small cluster of pixels experiencing a jump.

\begin{figure}
    \centering
    \includegraphics{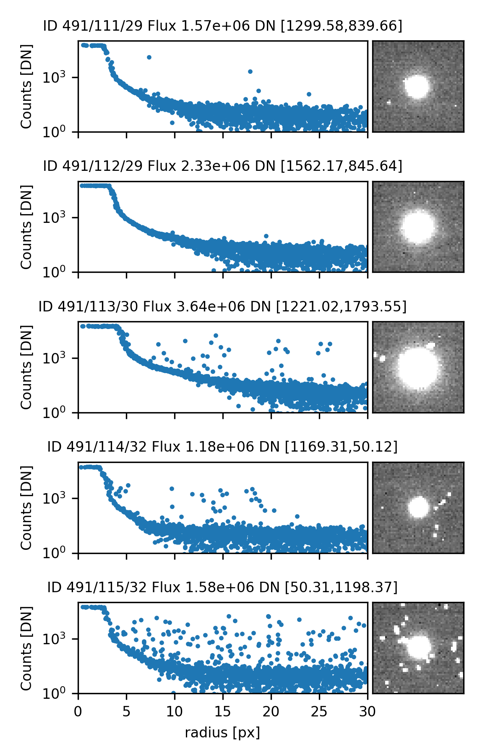}
    \includegraphics[]{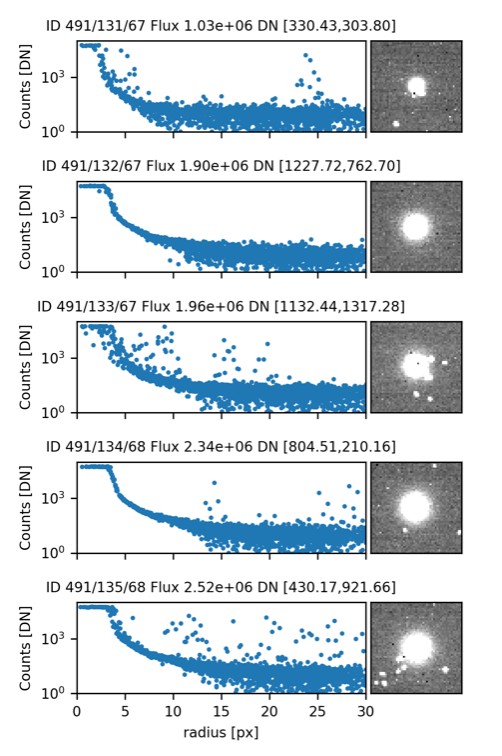}
    \caption{Examples of ``snowballs''. The image inlays show the correlated double sample frame difference where the event occurred with a size of 50 by 50 pixels and a grey scale of +/- 80 DN. The scatter plots show the corresponding radial profile around the center of the snowball, with the flat ``top-hat'' like curve towards small radii indicated the fully saturated region.}
    \label{fig:snowballs}
\end{figure}

One surprise during commissioning was the ubiquitous appearance of so-called ``snowballs'': up-the-ramp jump events with mostly (but not exclusively) a spherical region of heavily saturated pixels (the core, typical radius 2 to 5 detector pixels), plus a more extended region of elevated signal that steeply drops with increasing distance from the center of the saturated region. Many snowballs are also associated with a shower of more compact or ``classical'' cosmic ray events in the same CDS difference. A few examples for snowballs and their radial profiles are shown in Fig.~\ref{fig:snowballs}.

As of this writing the origin of the snowballs is not understood, but given the number of fully saturated pixels and the extended halo it is clear that high energies of many keV or even MeV must be involved.

\section{Conclusion}

As demonstrated during commissioning and presented in this paper, the NIRSpec detector system is operating and performing to predictions. Its read noise is consistent with the last on-ground measurements, indicating that the detector tuning has not changed. The total noise is higher than measured on-ground due to the impact of cosmic rays, but it is in line with the predictions and noise model used by the ETC. Combining this with the excellent throughput of the instrument\cite{GiardinoSPIE}, the sensitivity requirements for NIRSpec are expected to be met with margin.

\appendix    

\acknowledgments 
 
We would like to acknowledge the hard work and dedication of the NIRSpec commissioning team, the NIRSpec science readiness team, and all people involved in the commissioning of JWST and NIRSpec. This work would not have been possible without them.

\bibliography{report} 
\bibliographystyle{spiebib} 

\end{document}